# Composition and reactivity on Pd/Au(111) surface: A first-principle study

D.W().Yuan[1,2] X.G. Gong[3] and Ruqian Wu[1]

1. *Department of Physics and Astronomy, University of California, Irvine, Irvine, CA 92697-4575*

2. *ICTS, Chinese Academy of Sciences, Beijing,100080, China*

3. *Surface Science Laboratory and Department of Physics, Fudan University, Shanghai-200433, China*

(Dated: Sep. 19, 2006)

Abstract

Atomic configurations of two or three Pd substituents on the Au(111) surface are investigated using the first-principles pseudopotential plane wave approach. Pd atoms are found to form second neighborhoods on PdAu(111). The Pd-d band becomes narrow and well below the Fermi level, very different from those in a Pd film or bulk Pd. Nevertheless, surface Pd atoms are still active and serve as independent attractive centers towards adsorbates. The special ensembles are important for catalysis applications because of their ability to confine reactants in a small region.

In order to design and fabricate excellent bimetallic structures for catalysis applications, it is crucial to find efficient ways to control their chemical properties by manipulating lattice strain, constituents and growth conditions[1,2,3,4,5,6,7,8]. It has been recently recognized that the ensemble effect, associated with particular distribution patterns of active constituents, may play a key role in promoting chemical reactions[2,6]. The unusually high reactivity and selectivity of the PdAu(001) bimetallic catalyst towards vinyl acetate (VA) synthesis, for instance, mainly stem from the presence of second neighbor Pd pairs on Au(001)[9,10] that provide appropriate distance between reactants. Although the presence of different ensembles can be inspected using the scanning tunneling microscope (STM), very few atomic resolution images have been reported. More often, the existence of different ensembles are probed with ethylene, CO and H, using various surface sensitive techniques such as infrared reflection absorption spectroscopy (IRAS) and temperature programmed desorption (TPD)[7,8]. For PdAu bimetallic surfaces, which are among the most important mixed-metal catalysts, it is well established that Au tends to segregate to the surface even at moderate temperatures[11] because of the difference in their surface energies. The morphology of PdAu surfaces strongly depends on the experimental condition and annealing history[12,13,14]. From STM images and infrared data, Behm and coworkers concluded that the critical ensemble for CO adsorption on PdAu(111) is Pd monomer. This conclusion was supported by the extensive work of Goodman's group; they found only one feature at 2087 cm$^{-1}$ in IRAS for CO/PdAu(111) if the substrate is annealed up to 800 K. However, the correlation between experimental features and surface configurations have not yet been clearly established, and theoretical studies are therefore required to attain comprehensive understandings. Furthermore, ensembles with second neighbor Pd are seldom separated from isolated monomers. Since reactants confined in a region smaller than 5 Å should have enormous opportunity to interact, the former may play a significant role in catalysis applications and deserves special attention.

In this Letter, we report results of energetic and electronic properties of the PdAu(111) bimetallic surfaces using state-of-the-art density functional approaches. It was found that the formation of second neighbor Pd ensembles is energetically more favorable, especially around subsurface Pd dopants. Each Pd in these ensembles behaves like an independent attractive center towards small foreign molecules such as CO. It is not a surprise that TPD and IRAS measurements detect single features for CO on the PdAu(111) surface annealed under high temperatures.

The calculations are performed in the framework of density functional theory (DFT), using the generalized gradient approximation (GGA) for the description of exchange-correlation interaction[15]. The effects of ionic cores are represented by ultra-soft pseudopotentials[16], as implemented in the Vienna *ab initio* Simulation Package (VASP)[17]. The Au(111) surface is modeled with a 5-layer Au slab and a 15 Å vacuum in between. In the lateral plane, we use a 4×4 supercell so as to mimic cases with lower Pd concentration and sparse CO adsorption. Plane waves with an energy cutoff of 350 eV are used to expand wave functions, while integrals in the reciprocal space are evaluated through summations over 5×5×1 k-points in the Monkhorst-Pack grids[18]. The two bottommost Au layers are frozen at their bulk positions, whereas all the other atoms are fully relaxed with a criterion that requires having the calculated atomic forces smaller than 0.03 eV/ Å on each ion. As known, GGA calculations overestimate the lattice sizes of Au and Pd by 2%. To circumvent the "artificial stresses" in structural optimization procedures, we use the theoretical lattice size for the bulk Au, $a_{Au}$=4.18 Å, in the lateral plane throughout the calculations.

To quantitatively represent the energetic and thermal stability of different Pd ensembles, we define and calculate their formation energies as

$$\Delta E_{Pd} = -[E_{PdAu} - E_{Au-slab} + N_{Pd}(E_{Au-bulk} - E_{Pd-atom})]/N_{Pd} \qquad (1)$$

Here, $E_{PdAu}$, $E_{Au-surf}$, $E_{Au-bulk}$ and $E_{Pd-atom}$ represent the total energies of the PdAu surface, clean Au(111) slab, bulk gold (per atom), and isolated Pd atom, respectively. $N_{Pd}$ is the number of Pd atoms in the unit cell. The data of $\Delta E_{Pd}$ for different ensembles on the PdAu(111) surface are presented in Fig. 1. It is obvious that ensembles consisting of first neighbor Pd pairs (e.g., 1b and 1d) are energetically unfavorable. For instance, ensembles 1b and 1c have the same set of atoms but the former is higher in energy by 0.10 eV (or 0.05 eV per atom). The energy difference between the other two comparable ensembles,

1d and 1e, is even larger, 0.26 eV. According to a rough estimation based on the standard Boltzmann distribution model, the population ratio between 1b and 1c is smaller than 5% at room temperature while the chances of having ensemble 1d is negligible. Our results explain the observation made by Maroun et al from the atomic scale STM images, where Pd atoms predominantly form "monomers". The absence of nearest Pd neighborhoods can be explained through a pair-interaction model. By using total energies of the bulk Au, bulk Pd, and PdAu alloy, we find that the energy associated with each Au-Au, Pd-Pd and Pd-Au bond is 0.249 eV, 0.307 eV and 0.317 eV, respectively. These numbers are transferable to surface systems and, indeed, produce the formation energies in Fig. 1 surprisingly well[19], i.e., 3.56 eV for ensemble 1b and 3.52 eV for ensemble 1d. Clearly, the key factor that makes the nearest Pd neighbors unfavorable in PdAu systems is the high strength of PdAu bonds. This same reason was used to explain the distribution of Pd on PdCu surfaces deposited on Ru(0001)[20] as well as in ordered and disordered of bulk alloys[21].

Note that we find the formation of second neighbor Pd pairs somewhat favorable. For example, ensemble 1e is 0.03 eV more stable than three isolated monomers and their population ratio should be close to unity at room temperature (~ e/3). The energy As a matter of fact, the distribution of Pd in the atomic scale STM images for PdAu(111) is non-uniform and ensembles 1c and 1e can be found in several places. This tendency is much more evident on PdAu(001) since $\Delta E_{Pd}$ of ensembles comprising two or three Pd second neighborhoods are larger than that of an isolated Pd monomer on Au(001) by as much as 0.05-0.07 eV per Pd atom. Since the second neighbor distances on Au(111) and Au(001) are only 4.99 Å and 4.08 Å, not much larger than the sizes of many reactants, it is essential to discuss second neighbor ensembles separately from isolated Pd monomers. The most important aspect of the existence of second neighbor ensembles is their ability to confine several reactants in a small range for an extended duration, crucial for their subsequent reaction.

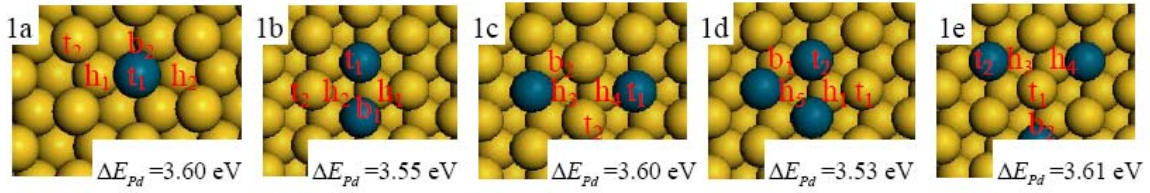

*FIG. 1: The calculated formation energies of selected ensembles with two and three Pd dopants in the surface layer of PdAu(111).*

Since most Pd atoms segregate into the interior region, it is important to investigate their influence on the formation of surface ensembles, particularly from the subsurface dopants. We define the formation energy of each surface atom as

$$\Delta E_{Pd}^s = -[E_{PdAu} - E_{Au+Pd(sub)} + N_{Pd}^s(E_{Au-bulk} - E_{Pd-atom})]/N_{Pd}^s \qquad (2)$$

Where $E_{Au+Pd(sub)}$ is the total energy of Au slab with one Pd atom in the subsurface layer, and $N_{Pd}^s$ is the number of Pd atoms in the surface layer. As displayed in Fig. 2, our data again indicates that first Pd neighbors between surface and subsurface atoms are unfavorable. However, the presence of subsurface Pd further promotes the formation of second neighbor Pd dimers, as manifested by values of $\Delta E_{Pd}^s$ for ensembles 2e and 2f.

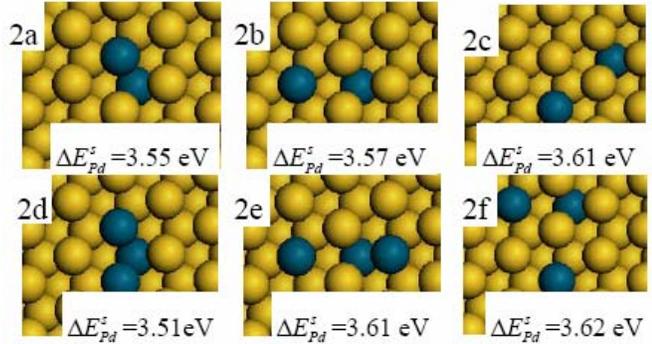

*FIG. 2: Calculated formation energies of selected ensembles with one Pd atom in the subsurface layer of PdAu(111).*

To understand why the Pd-Au bond is stronger than the Pd-Pd bond and also to access the chemical properties of PdAu(111), the projected density of d-states (PDOS) are plotted in Fig. 3 for the Pd atom and the adjacent Au in ensembles 1a and 1c. For comparison, results of the clean Au(111) and Pd(111) surfaces are also presented. It is interesting that the Pd-d band becomes very narrow, somehow similar to those of noble metals. In contrast, the PDOS of Au-d band remains almost unaffected from that of the clean Au(111) surface. Furthermore, direct hybridization between second neighbor Pd atoms is negligible since the PDOS curves for Pd in ensembles 1a and 1c overlap with each other over the entire energy range. From the charge density difference shown in the inset, it is clear that the effect of a Pd dopant is limited to its first neighbors. Pd-d shell strongly attracts electrons from the surrounding Au atoms and, meanwhile, draws electrons from its own s-band compared to that under conditions on Pd surfaces or in bulk Pd. As a result, the Pd-d band is essentially fully occupied in PdAu(111) and the atomistic features are somewhat restored. Despite the weak hybridization, Pd-Au bond encompasses significant ionic features and is stronger than the covalent/metallic Pd-Pd bond.

From the PDOS alone, one may perceive that Pd should not be much more active than Cu towards adsorbates. Meanwhile, Au sites are still inactive because of the negligible change in Au-PDOS from Au(111) to PdAu(111). To examine the local chemical properties, it is useful to calculate the chemisorption energy and site preference of various small molecules such as CO. In Table 1, we present the calculated adsorption energies, $E_{ad}$, and stretching frequencies, f, of CO on different ensembles of PdAu(111). As known, the values of $E_{ad}$ for molecule-metal systems are usually overestimated in DFT-GGA calculations [22]. The site preference is reliable only when energy differences are sufficiently large, a condition that is conveniently satisfied for CO on Pd(111) and PdAu(111). As shown in Table 1, CO prefers the fcc hollow site and $E_{ad}$ is 2.06 eV/molecule, about 0.5 eV larger than the experimental data, $E_{ad}$=1.47-1.53 eV [23]. Calculations using the revised Perdew-Burke-Ernzerhof (RPBE) GGA functionals [24] produce smaller $E_{ad}$ (data in brackets), by 0.24-0.28 eV, yet reserve the sequence of site preference. RPBE also reduces the adsorption energies for CO on Au(111) by the same amount and gives negative $E_{ad}$. This suggests that CO does not bind to a flat Au(111), which is in good agreement with experimental observations. Since CO adsorption energy above Au sites on PdAu(111) is not much larger than that on the clean Au(111) through GGA-PW91 calculations, it is anticipated that CO molecules only take the Pd sites while remaining highly mobile elsewhere on PdAu(111).

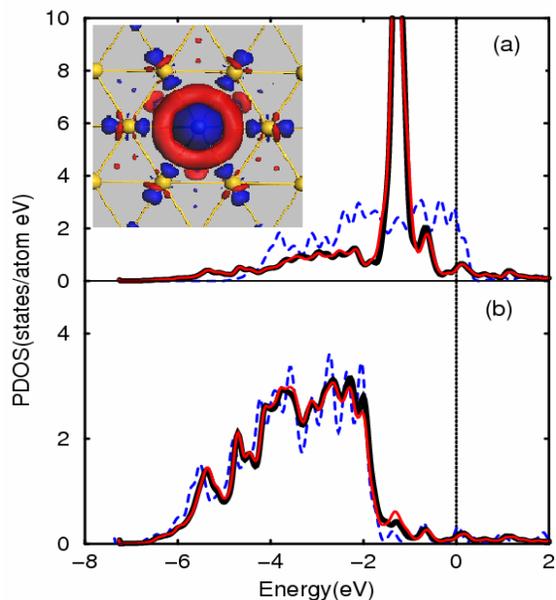

Fig. 3. The projected density of d-states of Pd (in panel a) and adjacent Au (in panel b) in the PdAu(111) surface with ensembles displayed in Fig. 1a (bold/black lines) and 1c (thin/red lines). Dashed/blue lines are PDOS of Au and Pd clean surfaces. Inset shows the charge difference $\Delta\rho = \rho_{PdAu} - \rho_{Au} - \rho_{Pd\_atom}$. The red and blue regions represent charge accumulation and depletion, respectively.

TABLE 1: The GGA-PW91 results of adsorption energies, $E_{ad}$ (in eV), and stretching frequency, $f$ (in $cm^{-1}$) of CO adsorbate on different surfaces. The abbreviations of adsorption sites are marked in Fig. 1 for different ensembles. Results in brackets are produced with the RPBE functionals.

| System | Site | $E_{ad}$(PW91) | f |
|---|---|---|---|
| Pd(111) | atop | 1.39[1.15] | 2040 |
|  | fcc-hollow | 2.06[1.80] | 1766 |
|  | hcp-hollow | 2.04[1.76] | 1770 |
|  | bridge | 1.85[1.57] | 1854 |
| Au(111) | atop | 0.22[-0.04] | 2048 |
|  | fcc-hollow | 0.20[-0.04] | 1819 |
|  | hcp-hollow | 0.21[-0.07] | 1838 |
|  | bridge | 0.25[0.01] | 1895 |
| PdAu(111)-1a | $t_1$ | 1.11 | 2056 |
|  | $t_2$ | 0.28 | 2068 |
|  | $h_1$ | 0.78 | 1854 |
|  | $h_2$ | 0.79 | 1815 |
|  | $b_2$ | 0.82 | 1895 |
| PdAu(111)-1b | $t_1$ | 1.23 | 2046 |
|  | $t_2$ | 0.32 | 2069 |
|  | $h_1$ | 1.41 | 1809 |
|  | $h_2$ | 1.43 | 1800 |
|  | $b_1$ | 1.54 | 1861 |
| PdAu(111)-1c | $t_1$ | 1.17 | 2043 |
|  | $t_2$ | 0.30 | 2045 |
|  | $h_3$ | 0.91 | 1806 |
|  | $h_4$ | 0.88 | 1800 |
|  | $b_2$ | 0.96 | 1907 |
| PdAu(111)-1d | $t_1$ | 1.25 | 2050 |
|  | $t_2$ | 0.35 | 2067 |
|  | $h_1$ | 1.46 | 1807 |
|  | $h_5$ | 1.93 | 1774 |
|  | $b_1$ | 1.67 | 1868 |
| PdAu(111)-1e | $t_1$ | 1.16 | 2049 |
|  | $t_2$ | 0.31 | 2070 |
|  | $h_3$ | 0.92 | 1781 |
|  | $h_4$ | 0.90 | 1824 |
|  | $b_2$ | 0.94 | 1889 |

On the preferential ensembles 1a, 1c, and 1e of PdAu(111), CO strongly prefers the $t_1$ (on Pd, cf. the site notations in Fig. 1) site. This site preference should not be changed by corrections to GGA since the energy differences are larger than 0.21 eV/molecule. Actually, inclusion of corrections typically reduces the value $E_{ad}$ much more for adsorptions on high-coordinate sites (e.g., the hollow site) and therefore should further enhance the energy preference on the $t_1$ site. Since 1a, 1c, and 1e are major ensembles on PdAu(111), one should observe that CO binds primarily to the top sites. This agrees well with the results found by Behm's and Goodman's groups[6,7]. The calculated CO stretching frequencies on ensemble 1a, 1c and 1e are 2043-2056 cm$^{-1}$, as compared to experimental value of 2087 cm$^{-1}$ (the deviation of 40 cm$^{-1}$ is caused by GGA). Evidently, it is quite difficult to distinguish these three ensembles through TPS and IRAS measurements with CO adsorption since they share almost identical values for either $E_{ad}$ or f. To this end, larger molecules such as ethylene and vinyl acid might be used to probe the effects of second neighbor ensembles, but the explanation of experimental data is expected to be more intricate. Our separate calculations indicate that $C_2H_4$ also prefers the atop site on PdAu(001) but the addition of a second neighbor Pd increases the adsorption energy from 0.56 eV to 0.69 eV.

It is worth mentioning that CO takes the three-fold hollow site only on the least favorable $Pd_3$ ensemble (i.e., 1d), and that the adsorption energy is large 1.93 eV. On the $Pd_2Au$ ensemble (i.e. 1b), CO prefers the bridge site between Pd atoms and $E_{ad}$ is 1.54 eV. As a result, the values f (1774 cm$^{-1}$ and 1861 cm$^{-1}$, respectively) are significantly reduced in these geometries since CO bond strength is much weakened through hybridization with Pd atoms underneath. Experimentally, two features are observed in the IRAS, 2087 cm$^{-1}$ and 1940 cm$^{-1}$, for CO adsorbed on a 5 ML Pd/5 ML Au surface if the surface is annealed to 600 K only. The second feature is close to the calculated frequency for CO on ensemble 1b, 1861 cm$^{-1}$. However, this feature disappears if the surface is annealed up to 800 K, which can be attributed to the removal of first Pd neighborhoods.

Finally, we want explore the mechanism of CO-PdAu(111) interaction. From the PDOS curves for CO/PdAu(111) in Fig. 4, one can see that Pd-d states above -1 eV are hardly disturbed by CO. This excludes the possibility of having major charge transfer between CO and PdAu(111). The main change in the CO side is the remarkable broadening of the $2\pi^*$ peak. Meanwhile, noticeable shifts are found for Pd states, from the main peak at -1.6 eV to the low energy regime. The resonant features suggest that the CO and Pd interact chiefly through hybridization between the CO-$2\pi^*$ orbital and Pd-d states. From the charge density difference displayed in the inset, one can find that electrons deplete from CO-5σ and Pd-$d_{z^2}$ states but fill the CO-2π* and Pd-$d_{xy}$ states. Apart from the exceptional charge accumulation (red region) right above Pd, the renown mechanism of CO-metal interaction, namely via donation from the CO-5σ state to metal and back donation from metal to the CO-2π* state, seems still applicable for CO/PdAu(111). The high activity of Pd in PdAu(111) results from the fact that Pd is ionic and has more electrons to share with adsorbates.

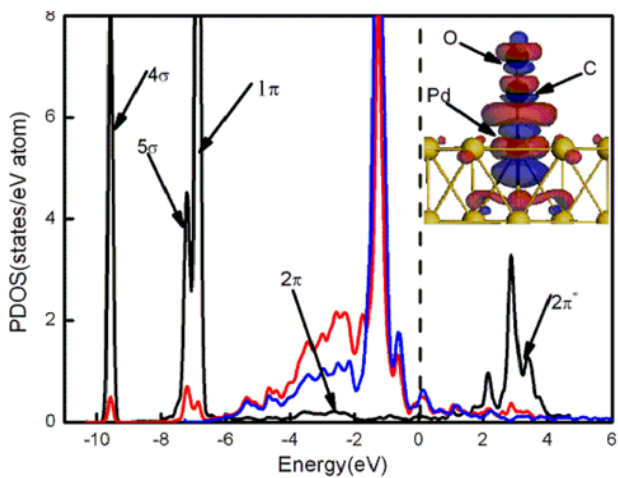

FIG. 4: The PDOS of CO (bold/black lines) and Pd (thin/red line) in CO/PdAu(111) on the $t_1$ site, accompanied by the PDOS of Pd (blue line) in the clean PdAu(111) surface. The inset displays the charge redistribution $\Delta\rho = \rho_{CO/Pd+Au} - \rho_{CO} + \rho_{Pd+Au}$. The red and blue regions represent charge accumulation and depletion, respectively.

In summary, using the first principles pseudopotential plane wave method, we studied the energetic stability of various ensembles on the PdAu(111) bimetallic surface. Our results explained the absence of nearest Pd neighborhoods and, significantly, reveal the existence of second neighbor ensembles. The Pd atoms are strong attractive centers for foreign adsorbates such as CO, and the second neighbor ensembles may trap several reactants in a small region, which is necessary for imminent chemical reactions. Future research in this direction may bring about efficient approaches towards the rational search and design of bimetallic nanocatalysts.

Work was supported by the DOE-BES (grant No: DE-FG02-04ER15611). RW acknowledges the help of Prof. D.W. Goodman for simulative discussions. DY was also supported by the ICTS, Chinese Academy of Science. XG was supported by the NSF of China, the national program for the basic research and research program of Shanghai. Calculations are performed on supercomputers in the NERSC.